\documentstyle[thmsa]{article}

\begin{document}

\title{Losses for microwave transmission in metamaterials for producing left-handed
materials: The strip wires. }
\author{E.V.Ponizovskaya$^1$ , M.Nieto-Vesperinas$^2$ and N.Garcia$^1$ \\
\ $^1$ Laboratorio de F\'{i}sica de Sistemas Peque\~{n}os y
Nanotecnolog\'{i}a,\\
C.S.I.C., c/Serrano 144, 28006 Madrid\\
$^2$ Instituto de Ciencia de Materiales de Madrid, Consejo\\
Superior de Investigaciones Cient\`{i}ficas, \\
Campus de Cantoblanco Madrid 28049, Spain}
\date{\today}
\maketitle

\begin{abstract}
This paper shows that the effective dielectric permitivity for the
metamaterials used so far to obtain left-handed materials, with strip wires
0.003cm thick, is dominated by the imaginary part at 10.6- 11.5 GHz
frequencies, where the band pass filter is, and therefore there is not
propagation and the wave is inhomogeneous inside the medium. This is shown
from finite-differences time-domain calculations using the real permitivity
values for the Cu wires. For thicker wires the losses are reduced and the
negative part of the permitivity dominates. As the thickness of the wires is
critical for the realization of a good transparent left- handed material we
propose that the strip wires should have thickness of 0.07-0.1cm and the
split ring resonators 0.015-0.03cm thick.
\end{abstract}

The subject of left-handed materials (LHM) is at present a prominent field
in optics and physics, due to the intriguing possibility of performing
negative refraction [1]. Both claims using metamaterials [2], and disclaims
[3], of observation of a negative refraction index have been done. The
disclaims were based on the fact that losses are important in the structures
so far built to observe negative refraction, and therefore the
electromagnetic waves are inhomogeneous in those metamaterials. Hence, for
experiments with a wedge-shaped geometry, there is a problem of correctly
interpreting the transmission measurements due to the non uniform absorption
in such a sample.

This letter points that the losses, for transmission of waves in
metamaterials, are critically depending of the thickness of the wires and of
the permitivity of the wires used. This is done by performing
finite-differences time-domain (FDTD) simulations By using strip wires
thicker than those used in [2] of 0.003cm thick, one can still have a
negative effective refractive index in the structure, like claimed in [2],
but much smaller losses than in their structure employed in the experiment
of [2] obtaining a good transparent LHM.

To show our line of argument, we point out first an error in calculations by
Shelby et al [4]. More specifically Fig5 of [4] is off by ten orders of
magnitude. We also conclude that the value of the damping constant g of the
magnetic permeability and dielectric permitivity that sets attenuation of
the wave propagation, is not 1GHz at the band pass frequencies: 10.6-11.5
GHz of the experiments of [4]. In fact the authors of [4] argue ''To match
the measure attenuation of the propagation band we set  $\gamma =1GHz$,
suggesting this structure has relatively large losses''. The value of their
calculation is given from Fig.5 of ref.4 in Fig.1a. However this is in error
as we have checked. Our calculation in the homogeneous materials
approximation that the authors of ref.4 point out to be valid (and we agree
with this), by using Eq.(4) of our work [3], gives the result shown in Fig.
1b. This, as can be seen, is ten orders of magnitude smaller than that given
in Ref.4. There is not question that the correct result using the
permeabilities, permittivities and the parameters of Ref.4 is that of Fig.
1b. This corrects Fig. 5 of [4]. Fig. 1b also proves that the correctly
calculated transmitted intensities are extremely small, and below the
experimental detection threshold (-55dB). Therefore, this disagreement with
the experiments of [4] forces g to have to be much smaller than 1GHz as
reported there.

Interestingly, the same authors claim in the paper where they report to
observe negative refraction [2], that the value of $\gamma =0.01GHz$.
However, the metamaterial is the same as in [4]. So how can g be factor of
100 different between both publications, being the same material?. Further,
in another recent paper [5] the same authors claim $\gamma =2GHz$ from a
fitting through transfer matrix method simulations. This value does not fit,
however, the experiments of [4]. Nevertheless, the simulations of [5] do not
have much connection with the experiment of [2] and [4] because the proper
value for the permittivity of copper wires: $-2000+(10^{6}\div 10^{7})i$ is
not used in the structure of the metamaterial of those simulations. Neither
the proper thickness of the strip wires (d) or the split rings resonator
(SRR) of the experiment of [2] and [4]: (0.003cm) is employed in [5]. The
value used in [5] for the thickness is 0.025-0.033cm (ten times larger than
[2] and [4]). Also, whereas the experimental strip wires used are 0.03mm
thick, those of the simulation in [5] are 1mm. The reason is that, as they
argue in [5], they cannot calculate for the experimental wire size yet.

The above remarked confusion in the values of the damping constant g is due
to two facts: (i) the expression used for the permittivity in [4] has
nothing to do with reality, and should be ruled out as has brilliantly been
proved by a recent comment by Walser et al. in [6] (see also the full FDTD
calculations for the effective permitivity in Fig. 2), and (ii) estimations
from the wires thickness indicate that $\gamma \simeq 0.2-0.5GHz$. Thus,

for modelling metamaterial structures that behave like a left-handed
material, one should:

(1) Study the effective permitivity of the structure as a function of both
the frequency and of the strip wire thickness.

(2) Carry out a similar study for the behaviour of the split ring resonators.

(3) Combine the study for the strip wires and the resonators together.

We stress that the proper permitivity, and not approximations, of the
metallic elements is required. We next show how the thickness of the wires
is crucial to obtain a good transparent left-handed material.

Fig. 2(a) shows the effective permitivity versus frequency n of the incident
microwave, for an array of copper wire strips ($\epsilon _{Cu}=-2000+i10^{6}$
) of square section with size 0.003 cm, (size of unit cell 0.5cm ), obtained
through an FDTD calculation. This is the size of the strip wires of the
experiments of [2] and [4]. As seen, the imaginary part of e is up to 1.5
for $\nu <5Ghz$, and is positive, about 0.5 near the microwave frequencies
of interest around 11 Ghz . On the other hand, the real part of e is small
and practically zero for n around 11 Ghz. So that clearly ther permitivity
at 11GHz is dominated by its imaginary part and the real part is practically
zero.

The absorption of the transmitted wave, shown in Fig 2(b) is rather large:
-5dB near the frequency of the experiments, whereas at that same frequency
the transnmitivity is -15dB.. This is for three rows of cylinders as the
number of rows increases the transnmitivity decreases. On the other hand, if
the thickness of the wires is larger, 0.1 cm, as shown in Fig.3(a), the
effective permitivity e of the structure now has a practically zero
imaginary part, whereas its real part is now negative and large.
Accordingly, the corresponding absorption and transnmitivity are now -20 and
-60dB respectively (Fig.3(b)). Note however, that the important feature in
these curves is the absorption distribution, which is large for the thin
wires, and small for the thicker wires. Analogous things happen for m that
has a negative real part and zero imaginary part for the thicker SSRs, the
product em now becomes positive and the structure becomes dielectric with
low losses.

In other words, in a structure where the imaginary part of e is large, like
in the thin wires, the absorption will remain large for the whole wires plus
SSR structure, the SSRs also being thin, produce a m that now has a negative
real part and a not negligible imaginary part. Then the product em being a
complex number with a negative real part and a rather large negative
imaginary part, therefore the effective refractive index $n=(\epsilon \mu
)^{1/2}$ having therefore a large imaginary component. Whereas for the
thicker strip wires and thicker SSRs m has a negative real part and a
negligible imaginary part, and then the structure has an effective product
em that is positive and the refractive index $n=(\epsilon \mu )^{1/2}$ is a
negative real number, the material behaving now like a transparent
dielectric.

>From all these calculations we infer therefore that, in order to obtain a
structure with low absorption and thus a transparent LHM with an effective
refractive index that is practically real and negative, one should make an
array of SSRs and strip wires whose thickness are between 0.015- 0.03cm,
five times bigger than in [2] and [4] for the SSRs, and about 0.07 - 0.1cm
for the strip wires.

This work has been supported by the Spanish DGICyT.

REFERENCES.

1. V. G. Veselago, Sov. Phys. Usp. 10, 509 (1968).

2. R. A. Shelby, D. R. Smith and S. Schultz, Science 292, 77(2001).

3. N. Garcia and M. Nieto, Optics Lett. 27, 885 (2002).

4. R. A.Shelby, , D. R. Smith, S. C. Nemat-Nasser and S. Schultz, Appl.
Phys. Lett. 78, 489 (2001)

5. D. R. Smith, S. Schultz, P. Markos and C. M. Soukoulis, Phys. Rev. B65,
195103 (2002)

6. R. M. Walser, A. P. Valanju and P.M. Valanju, Phys. Rev. Lett. 87,
119701-1 (2001).

FIGURE CAPTIONS

Fig.1 (a) Calculation of ref. 4, Fig.5 using a value of $\gamma =1GHz$ and
the homogeneous medium approximation. This result is off by ten orders of
magnitude. The good calculation is presented in Fig. 1b.

Fig.2. (a) Real and imaginary part for the effective permitivity of a
squared array of $d=0.003cm$ thick strip wires of Cu as used in the
metamaterials [2,4]. The results are practically the same for two or three
rows of wires. The calculation is for s-polarization, i.e electric field
parallel to the wires. The imaginary part dominates and the real part is
practically zero at 121 GHz where the experiment shows the band pass. (b)
shows the FDTD calculations (dots and rhombus) for the reflectivity and
transnmitivity of the structure of wires. The lines are the fitting that
give the values of e in (a) using the homogenous medium approximation.

Fig.3(a) The same as in Fig.2a for wires $d=0.1cm$ thick. Now the real part
is negative and dominates over the imaginary part. These data are obtained
from the FDTD calculations in (b) by fitting the simulations data (dots and
rhombus) using the homogenous

\end{document}